\documentclass[twocolumn,showpacs,preprintnumbers,amsmath,amssymb]{revtex4}

\usepackage{graphicx}
\usepackage{dcolumn}
\usepackage{bm}

\begin{document}

\preprint{APS/xxxx}

\title{Response of Spiking Neurons to Correlated Inputs}

\author{ {\bf\sc Rub\'en Moreno, Jaime de la Rocha, Alfonso Renart}}
\altaffiliation[]{Present address: Center for Complex Systems, Brandeis University, Waltham, MA 02454, USA.}
\author{\bf\sc N\'estor Parga}
\thanks{To whom correspondence should be addressed.}
\affiliation{
  \it Departamento de F\'{\i}sica Te\'orica, 
  Universidad Aut\'onoma de Madrid, 
  Cantoblanco, 28049 Madrid, Spain.
  }

\begin{abstract}
The effect of a temporally correlated afferent current on the firing
rate of a leaky integrate-and-fire (LIF) neuron is studied. This
current is characterized in terms of rates, auto and
cross-correlations, and correlation time scale $\tau_c$ of excitatory
and inhibitory inputs. The output rate $\nu_{out}$ is calculated in
the Fokker-Planck (FP) formalism in the limit of both small and large
$\tau_c$ compared to the membrane time constant $\tau$ of the
neuron. By simulations we check the analytical results, provide an
interpolation valid for all $\tau_c$ and study the neuron's response
to rapid changes in the correlation magnitude.
\end{abstract}


\maketitle

One of the most fundamental questions in neuroscience is to understand
the way neurons communicate. There is growing evidence that temporal
correlations in the spike emission times
play a relevant role in the transmission of information (see,
e.g. \cite{Sal+01a}).  Although correlations are indeed present
throughout cortex \cite{Zoh+94,deC+96}, their functional role is
controversial \cite{Sha+98,Sof+93,Sin99R}. A relevant issue is how
temporal correlations in a population affect the response of a
postsynaptic neuron.  Most of the work in this direction has been
numerical, and little is known even for simple neuron models such as
the LIF neuron \cite{Sha+98,Sal+00}.  A better knowledge of how
correlations affect the neuron's input-output function would be
useful, for instance, to study networks of these neurons, where
correlations are unavoidable. A related issue is the speed with which
changes in the correlations of inputs can be detected by a
post-synaptic cell. In this letter we study both questions and
interpret our results in the context of experiments on auditory
processing \cite{deC+96}.  The main conclusions are: 1) the neuron's
output rate is sensitive only to precisely synchronized inputs
($\tau_c < \tau$); 2) the response decreases (increases) with the
timescale $\tau_c$ for positive (negative) correlations, and
increases (decreases) with their magnitude $\alpha$; 3) this increase
is larger for afferent currents in the fluctuation-dominated
(balanced) state than for those in the drift-dominated (unbalanced)
state; 4) the response increases until it reaches a saturation value
if the magnitude and time scale of the correlations are increased
simultaneously while keeping their ratio fixed; and 5) the neuron
response to sudden changes in the size of the correlations is very
fast, regardless of the magnitude of the change.\\


\vspace{-0.3cm}
\noindent
{\bf The neuron model and Input statistics.}  
The depolarization
membrane potential $V(t)$ of a LIF neuron evolves from the reset
voltage $H$ according to 

\begin{equation}
\dot{V}(t) = - \frac{V(t)}{\tau} + I(t) 
\label{IF_equation}
\end{equation}

\noindent
where $I(t)$ is the afferent and $\tau$ is the membrane time
constant \cite{Ric77}. When the input drives the potential to a threshold value
$\Theta$, a spike is emitted and the neuron is reset to $H$, from
where it continues integrating the signal after a refractory
time $\tau_{ref}$. The afferent current $I(t)$ is

\begin{equation}
I(t)=J_{E} \; \sum_{i=1}^{N_E} \sum_{k} \delta(t-t_{i}^{k}) - 
     J_{I} \; \sum_{j=1}^{N_I} \sum_{l} \delta(t-t_{j}^{l})
\label{eq:current}
\end{equation}

\noindent 
where $t_{i(j)}^{k(l)}$ represents the time of the $k$-th ($l$-th)
spike from the $i$-th excitatory ($j$-th inhibitory) pre-synaptic
neuron, and $N_{E(I)}$ and $J_{E(I)}$ respectively represent the
number of inputs and size of the post-synaptic potentials from the excitatory (inhibitory)
afferent populations. We work in the limit of infinitely fast post-synaptic
currents, in which these are represented by delta functions.
We consider stochastic spike trains with exponential autocorrelations
with time constant $\tau_c$

\begin{eqnarray}
C_{p}(t,t')&\equiv&< \sum_{k,k'}
\delta(t-t_{i}^{k})\delta(t'-t_{i}^{k'}) > -\nu_p^2 \nonumber \\ &=&
\nu_p \delta(t-t') + \nu_p \left(\frac{F_p-1}{2 \tau_{c}}\right) \;
e^{-\frac{\mid t-t'\mid}{\tau_c}}
\label{autocorrelation}
\end{eqnarray}

\noindent
Here $p=E,I$, and $\nu_p$ and $F_p$ are the firing rate and
Fano factors of the spike counts (for infinitely long time windows) 
of the individual trains from
population $p$. Notice that if $F_p=1$ spikes are uncorrelated
(Poisson process) and that for $F_p>1$ spikes are positively
correlated while for $F_p<1$ they are negatively correlated. A similar 
expression for the autocorrelation of individual spike trains has been
used in \cite{Bru+98b} in a study of the effect of synaptic
filters. This problem is technically different from ours because those
filters integrate out the Dirac delta in the correlation (see
eq.(\ref{eq:current_correlation}) below). We also consider 
exponential cross-correlations

\begin{eqnarray}
C_{pq}(t,t')&\equiv&< \sum_{k_p,k_q} \delta(t-t_{i}^{k_p})\delta(t'-t_{j}^{k_q}) >-\nu_p \nu_q
\nonumber \\
&=& \sqrt{\nu_p \nu_q}\left(\frac{\rho_{pq} \;\sqrt{F_p\; F_q}}{2 \tau_{c}} \right)\; e^{-\frac{\mid t-t'\mid}{\tau_c}}
\label{crosscorrelation}
\end{eqnarray}

\noindent
between the trains $(i,j)$ in populations $p$ and $q$ ($p,q=E,I$). The 
magnitude of the cross-correlations is determined by the 
correlation coefficients $\rho_{pq}$ of the spike counts. For simplicity, we
consider all correlations in the problem to have the same time 
constant $\tau_c$.
The reason why the Fano factors appear in eq. (\ref{crosscorrelation}) is that 
the time integral of the cross-correlation has to be zero 
if one of the trains does not have spike count fluctuations. 
The correlation of the total afferent current is:

\begin{eqnarray}
C(t,t') &\equiv& <(I(t)-<I(t)>)(I(t')-<I(t')>)>  
\nonumber \\
        &=& \sigma_w^2 \delta(t-t') + 
\frac{\Sigma_2}{2 \tau_c} \; e^{- \frac{\mid t-t' \mid}{\tau_c}} \;\;,
\label{eq:current_correlation}
\end{eqnarray}

\noindent
where $\sigma_w^2$ is a white noise variance and $\Sigma_2$ is the 
contribution to the total variance, $\sigma_{eff}^2=\sigma_w^2+\Sigma_2$,
arising from correlations in the input spike trains:

\begin{eqnarray}
\sigma_w^2 &=& J_E^2 N_E \nu_E + J_I^2 N_I \nu_I 
\nonumber \\
\Sigma_2 &=& J_E^2 N_E \nu_E [(F_E - 1) + 
f_{EE } (f_{EE }N_E-1) F_E \;\rho_{EE}]
\nonumber \\
& & \mbox{} + J_I^2 N_I \nu_I [(F_I - 1) + 
f_{II} (f_{II }N_I-1) F_I \;\rho_{II}]
\nonumber \\ 
& & \mbox{} - 2 \; J_E J_I \; f_{EI} f_{IE} N_E N_I \;  \sqrt{\nu_E \nu_I } \; \sqrt{F_E \; F_I}  \rho_{EI}
\label{eq:variances}
\end{eqnarray}

\noindent
We suppose that only a fraction of presynaptic neurons can be
correlated with each other. The four parameters $f_{pq}$ denote the
fraction of correlated neurons from populations $p$ and $q$. 
The input current $I(t)$ is assumed to be Gaussian, condition which
naturally holds when the neuron is receiving a large barrage of spikes
per second \cite{Ric77}, each one inducing a membrane depolarization
$J$ very small compared to the distance between the threshold and
reset potentials, i.e., qualitatively $\frac{J F}{(\Theta-H)}(1+f N
\rho )\ll 1$.  Thus, the input can be described in terms of the mean
$\mu= J_E N_E \nu_E - J_I N_I \nu_I$, the variance $\sigma_w^2$, the
parameter $k \equiv \sqrt{\tau_c/\tau}$, and the {\it correlation
magnitude} $\alpha \equiv \Sigma_2/\sigma_w^2$ \footnote{For a
renewal spike trains the Fano factors in the above equations are related to
the coefficients of variation of their inter-spike-intervals as
$F_p=CV_p^2$.}.


\vspace{0.5cm}
\noindent
{\bf The analytical solution.} We express the input current $I(t)$ as

\begin{eqnarray}
I(t) & = & \mu +  \sigma_w \eta(t)+ \sigma_w  \frac{\beta}{\sqrt{2\tau_c}}  z(t) 
\label{current} \\
\dot{z}(t) & =   & - \frac{z}{\tau_c} + \sqrt{\frac{2}{\tau_c}} \eta(t) 
\label{dynamic_z}
\end{eqnarray}

\noindent
where $\eta(t)$ is a white noise random process with unit variance,
$\beta=\sqrt{1+\alpha}-1$ and $z(t)$ is an auxiliary colored random
process which obeys eq.(\ref{dynamic_z}) with the same white input noise
$\eta(t)$. Using eqs.(\ref{current}-\ref{dynamic_z}) is easy to check
that $I(t)$ is exponentially correlated in the stationary regime, with
correlations that read exactly as (\ref{eq:current_correlation}).



Associated to the stochastic diffusion process defined by eqs.
(\ref{IF_equation}, \ref{current}, \ref{dynamic_z}), we 
have the stationary FP equation\cite{Ris89}

\begin{equation}
[L_x  + \frac{L_z}{k^2} 
           + \frac{2}{k} \frac{\partial}{\partial x}(
                \frac{\partial}{\partial z} -\frac{\beta z}{2})]f
= - \tau \delta(x- \sqrt{2} \hat{H}) J(z)
\label{FP-equation}
\end{equation}

\noindent
where  $L_u = \frac{\partial }{\partial u} u +
\frac{\partial^{2} }{\partial^{2} u}$.
Besides, $V=\mu \tau + \sigma_w \sqrt{\frac{\tau}{2}} x$, 
$\hat H = \frac{H-\mu \tau} { \sigma_w \sqrt{\tau}}$ and
$\hat \Theta = \frac{\Theta-\mu \tau} { \sigma_w \sqrt{\tau}}$.

The function $f(x,z)$ is the steady state probability 
density of having the neuron in the state $(x,z)$.  
The key quantity $J(z)$ is the escape
probability current. It appears in eq.(\ref{FP-equation})
as a source term representing the reset effect: whenever the potential
$V$ reaches the threshold $\Theta$, it is reset to the value $H$ with
a distribution in $z$ that is unknown. The particular distribution of
$z$ will depend on the value of $\tau_{ref}$.
The escape current must be determined consistently using 
the normalization of the probability density,
$\tau_{ref}\nu_{out} + \int_{-\infty}^{\hat \Theta} 
dx \int_{-\infty}^{\infty} dz f(x,z)=1$, and
the threshold vanishing condition, $f( \sqrt{2} \hat \Theta,z)=0$.
The output firing rate is given by $\nu_{out}=
\int_{-\infty}^{\infty} dz J(z)$.


{\it Small $\tau_c$ expansion ($\tau_c \ll \tau $)}. In this regime the
quantities $k$ and $\alpha$ are treated as perturbative parameters. If
we suppose that the correlation time $\tau_c$ is very small compared
to the refractory time $\tau_{ref}$ ($\tau_{ref}\gg\tau_c $), the
escape current can be written as $J(z)= \nu_{out} e^{-z^2/2}/\sqrt{2
\pi}$ \cite{Bru+98b}.  We find $\nu_{out}$ analytically by expanding 
eq. (\ref{FP-equation})
in powers of $k=\sqrt{\tau_c/\tau}$, and calculating the terms exactly
for all $\alpha=\Sigma_2/\sigma_w^2$ for the zero order, and perturbatively in $\alpha
\geq 0$ up to the first non trivial correction for the first
order. The obtained firing rate can be written as

\begin{equation}
\nu_{out}= \nu_{eff} - \alpha \sqrt{\tau_c \tau}  
                \nu_0^2  R(\hat \Theta) \;\;
\label{nu_out_small_tauc}
\end{equation}

\noindent
Here $R(t)= \sqrt{\frac{\pi}{2}} e^{t^2}(1 + \rm{erf}(t))$, where 
$\rm{erf}(t)$ is the {\it error function}, and 
the rates $\nu_{eff}$ and $\nu_0$  are defined as

\begin{eqnarray}
\nu_{eff}^{-1}  &= & \tau_{ref} + \sqrt{\pi} \tau 
\int_{\hat H_{eff}}^
{\hat \Theta_{eff}} dt e^{t^2} (1 + \rm{erf}(t))  
\nonumber \\
\nu_0^{-1} &=  &  \tau_{ref} +  \sqrt{\pi} \tau 
\int_{\hat H}^
{\hat \Theta} dt e^{t^2} (1 + \rm{erf}(t)) \;\;
\end{eqnarray}

\noindent
The effective reset and threshold are defined as $\hat \Theta_{eff}=
\frac{\Theta-\mu\tau}{\sigma_{eff} \sqrt{\tau}}$ and $\hat H_{eff}=
\frac{H-\mu \tau}{\sigma_{eff} \sqrt{\tau}}$.  $\nu_0$ is the mean
firing rate of a LIF neuron driven by white noise \cite{Ric77}. 
Hence, eq.(\ref{nu_out_small_tauc}) implies that
when $\tau_c=0$ the problem is equivalent to considering 
an uncorrelated input with an effective signal variance
$\sigma_{eff}^2=\sigma_w^2+\Sigma_2$. In this case, our solution is
{\em exact} for all $\alpha$. When $\tau_c
\neq 0$, the expression is only correct for small values of both $k$
and $\alpha
\geq 0$. Here the analytical result applies only when $\alpha \geq 0$,
but we checked by numerical simulations that the same formula for the
output rate is also valid for $\alpha < 0$.


{\it Large $\tau_c$ expansion ($\tau_c \gg \tau $)}. In this limit the perturbative
parameter is $k^{-1}$. Now the escape probability current $J(z)$ must
be derived from the FP eq. (\ref{FP-equation}). If we assume that
$\tau_{ref}\ll\tau_c$,  

\begin{equation}
J(z)= -\frac{1}{\tau} \frac{\partial}{\partial x} 
           f(x,z)|_{x= \sqrt{2} \hat \Theta}
\label{eq:escape_probability}
\end{equation}

\noindent
This expression generates an additional constraint that should hold in
addition to the conditions defined above.  Using standard perturbative
techniques we find $J(z)$ and the mean firing rate, 
up to $O(k^{-2})$:

\begin{eqnarray}
&&J(z) = \frac{e^{-z^2/2}}{\sqrt{2 \pi}} [ \nu_0 +
   \sqrt{\frac{\tau}{\tau_c}}
      \frac{(2 + \beta) \nu_0^2 (R(\hat \Theta) -
      R(\hat H))}{1-\nu_0 \tau_{ref}} z
\nonumber \\
&& \;\;\;\;\;\;\;\;\;\;\;\;   + \frac{C}{\tau_c}  +
 \frac{(2 + \beta)C}{\beta \tau_c (1-\nu_0 \tau_{ref})}
        (z^2-1) ]
\label{expression-current-equation} \\
&&\nu_{out} = \nu_0 + \frac{C}{\tau_c}
\label{nu_out_big_tauc} \\
&&C  \equiv \alpha \tau^2 \nu_0^2 
[ \frac{\tau \nu_0 (R(\hat \Theta)-R(\hat H))^2}{1-\nu_0 \tau_{ref}} 
- \frac{\hat \Theta R(\hat \Theta) - \hat H R(\hat H)}{\sqrt{2}}] 
\nonumber
\end{eqnarray}

\noindent
Note that $\nu_{out}$ converges to $\nu_0$ when $\tau_c \gg \tau $.

\begin{figure}
\includegraphics[width=8cm,height=4.3cm,angle=0]{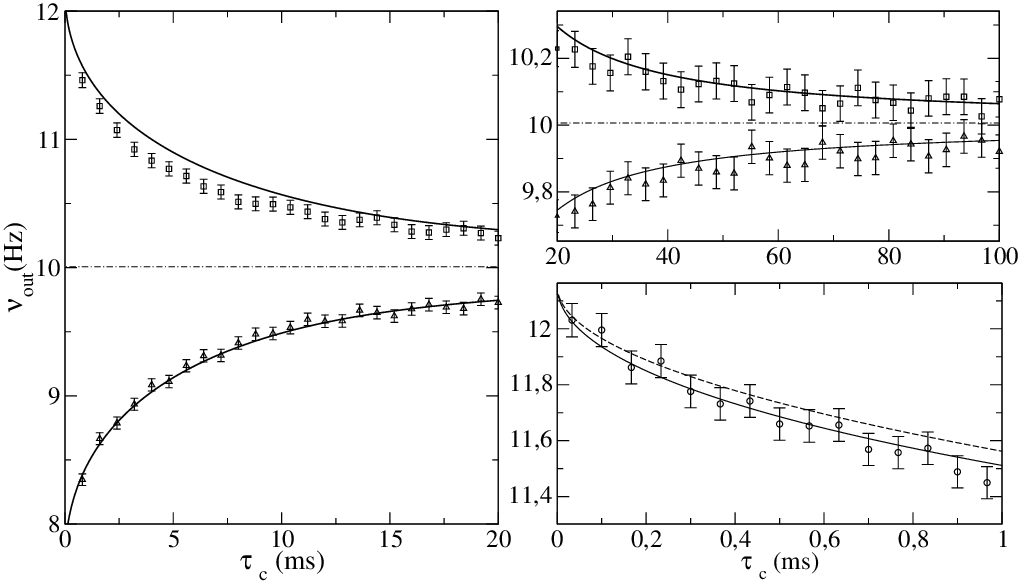}
\caption{\label{fig:grafica1}
Theoretical predictions and simulation results for
$\nu_{out}$ as a function of $\tau_c$.  {\bf Left}: $\alpha=0.21$
(upper curve) and $\alpha=-0.19$ (lower curve). {\bf Top right}: same
but for larger values of $\tau_c$. {\bf Bottom right}: the case
$\alpha=0.21$ for very small values of $\tau_c$.  {\em Full lines}:
interpolations between the small and large $\tau_c$ theoretical
predictions performed at the interpolating time $\tau_{c,inter}=14ms$.
{\em Dashed line}: small $\tau_c$ predictions from
eq.(\ref{nu_out_small_tauc}). {\em Horizontal line}: response to
white noise activation ($\alpha=0$).  Other parameters are
$\tau=10ms$, $\tau_{ref}=0ms$, $\Theta=1$ (in arbitrary units), $H=0$,
$\mu=81.7s^{-1}$, $\sigma^2_w=2.1s^{-1}$.  Although the small $\tau_c$
expansion requires $\tau_{ref} \neq 0$ the simulation shows that this
prediction is good even for zero $\tau_{ref}$.  
}
\end{figure}

\vspace{0.5cm}
\noindent
{\bf Results and comparison with numerical simulations.}  
We have performed numerical simulations of a LIF neuron driven by
Gaussian exponentially correlated input using
eqs. (\ref{IF_equation},\ref{current},\ref{dynamic_z}) with a twofold
motivation. First, they can be used to check the analytical results
given in eqs.(\ref{nu_out_small_tauc},\ref{nu_out_big_tauc}) and,
second, thet can be employed to determine higher order terms in the
perturbative expansions by interpolating the output rate between the
regimes of low and high $\tau_c$. The interpolating curves have been
determined by setting the firing rate in the small correlation time
range ($\tau_c < \tau$) as $\nu_{out}=\nu_{eff}+A_1 \sqrt{\tau_c}+ A_2
\tau_c$ where $A_1$ and $A_2$ are unknown functions of $\alpha$ and of
the neuron and input parameters, while in the large correlation time
limit ($\tau_c > \tau$) the expression given in
eq.(\ref{nu_out_big_tauc}), $\nu_{out}=\nu_0+C/\tau_c$, was used. The
functions $A_1$ and $A_2$ are determined by interpolating these two
expressions with conditions of continuity and derivability at a
convenient interpolation point $\tau_{c,inter} \sim \tau$.  Although
we have calculated analytically the function $A_1$
(eq.\ref{nu_out_small_tauc}) for small $\alpha$, this procedure takes
into account higher order corrections which match more accurately the
observed data for larger values of $\alpha$. Fig. \ref{fig:grafica1}
shows an example the good agreement between theory and simulations.
When positive correlations are considered ($\alpha > 0$), the
interpolation procedure is robust to changes in $\mu$ and
$\sigma_w^2$. For negative correlations, changing these parameters
sometimes results in lower quality fits. In these cases we have added
to the expansion in eq.(\ref{nu_out_big_tauc}) an extra term, so that
$\nu_{out}=\nu_0+C/\tau_c+B_1/\tau_c^2$. This is used to match at
$\tau_{c,inter}$ the small $\tau_c$ regime which is set as
$\nu_{out}=\nu_{eff}+B_2 \sqrt{\tau_c}$.

As it can be appreciated in Fig. \ref{fig:grafica1}, the response
increases as $\tau_c$ decreases (at fixed positive $\alpha$). This
corresponds to the intuitive result that positive correlations between
the pre-synaptic events produce a larger enhancement in the output
firing rate as the temporal window over which they occur decreases.
We have also considered a situation where the correlation magnitude
increases with $\tau_c$ as $\alpha = \gamma \tau_c$, for a fixed
$\gamma > 0$.  Eqs. (\ref{nu_out_small_tauc}) and
(\ref{nu_out_big_tauc}) suggest that the rate increases and saturates
as a function of $\tau_c$, because it depends only on the ratio
$\alpha / \tau_c$ in the long $\tau_c$ limit. We checked this
conclusion with simulations using the same parameters as in Fig.
\ref{fig:grafica1} (data not shown).
Note, however, that this manipulation
does not isolate the effect of changing the temporal range of the 
correlations, since now $\alpha$, which depends on the pre-synaptic
rates, Fano factors, etc, has to increase linearly with $\tau_c$.

\begin{figure}
\includegraphics[width=8cm,height=4.3cm,angle=0]{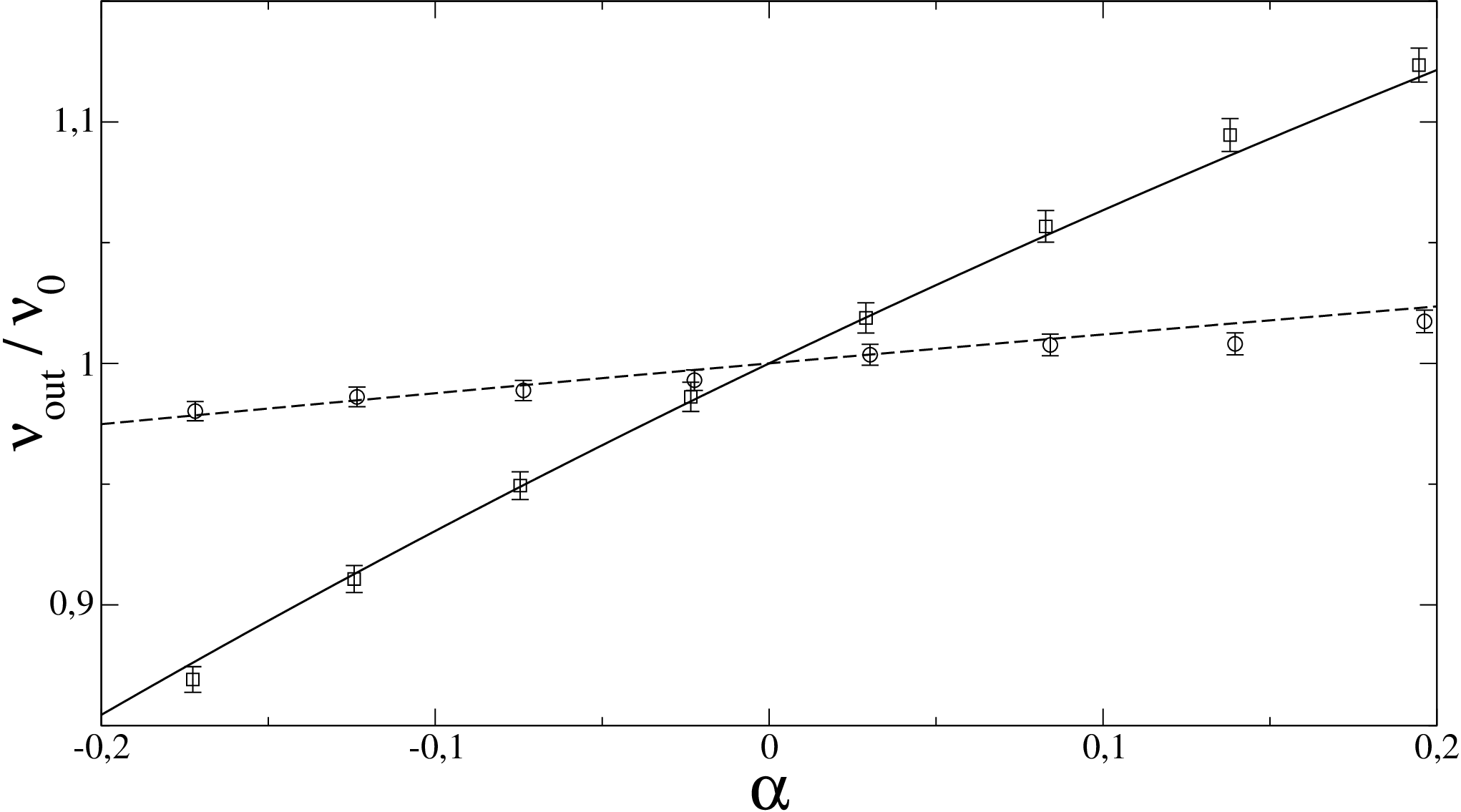}
\caption{\label{fig:grafica2}
Theoretical predictions and simulation results for
$\nu_{out}/\nu_0$ as a function of $\alpha$ for the balanced and the
unbalanced states. The neuron is much more sensitive to $\alpha$ in
the balanced regime (full line, $\mu=40s^{-1}$) 
than in the unbalanced regime (dashed line, $\mu=110s^{-1}$). 
In both cases $\sigma^2_w=30s^{-1}$, $\tau_c=1 ms$ 
and the other parameters are
as in Fig. \ref{fig:grafica1}. With these parameters $\nu_0=16.9Hz$
in the balanced state and $\nu_0=69.5Hz$ in the unbalanced state.}
\end{figure}

At fixed $\tau_c$, the rate increases with $\alpha$, as shown in
Fig. \ref{fig:grafica2}. The mean current, $\mu$, and the white
noise variance, $\sigma^2_w$, have been chosen so that the afferent
current puts the neuron either in the fluctuation-dominated or in the
drift-dominated regime \cite{Sha+98}.  Notice that the response is
more sensitive to changes in the correlation magnitude ($\alpha$) in
the balanced than in the unbalanced state,
in agreement with the findings in \cite{Sal+00} for similar neuron models.


We can also infer how fast a LIF neuron responds to
changes in the correlation magnitude $\alpha$ at {\em fixed} afferent
mean current and white noise variance $\sigma_w^2$.  
It is easy to verify that the instantaneous rate for
the time dependent FP equation can be expressed as 
(for the sake of clarity we have come back to the physical quantity
$V$ and used its distribution $P(V,z,t)$)

\begin{equation}
\nu_{out}(t) = - \frac{\sigma_w^2 (t)}{2} \frac{\partial
}{\partial V} \int_{-\infty}^{\infty} dz P(V,z,t)|_{V=\Theta}
\label{eq:rate_t}
\end{equation} 

\noindent
As we have seen, the exact solution for
$\tau_c=0$ corresponds to a renormalization of $\sigma_w^2$ to
$\sigma_{eff}^2$. This gives $\nu_{out}(t) = - \frac{\sigma_{eff}^2
(t)}{2} \frac{\partial}{\partial V} \int dz P(V,z,t)|_{V=\Theta}$. Now
it is clear that any change in $\sigma_{eff}^2$ will produce an
immediate change in $\nu_{out}$ \cite{Sil+02}. This means that when
$\tau_c=0$, changes in both correlation magnitude ($\alpha$) and white
noise variance ($\sigma_w^2$) will be felt immediately by the firing
response of the neuron. By analyticity arguments, the response under
changes in $\alpha$ will be also fast for non-zero, small $\tau_c$.

\begin{figure}
\includegraphics[width=8cm,height=4.3cm,angle=0]{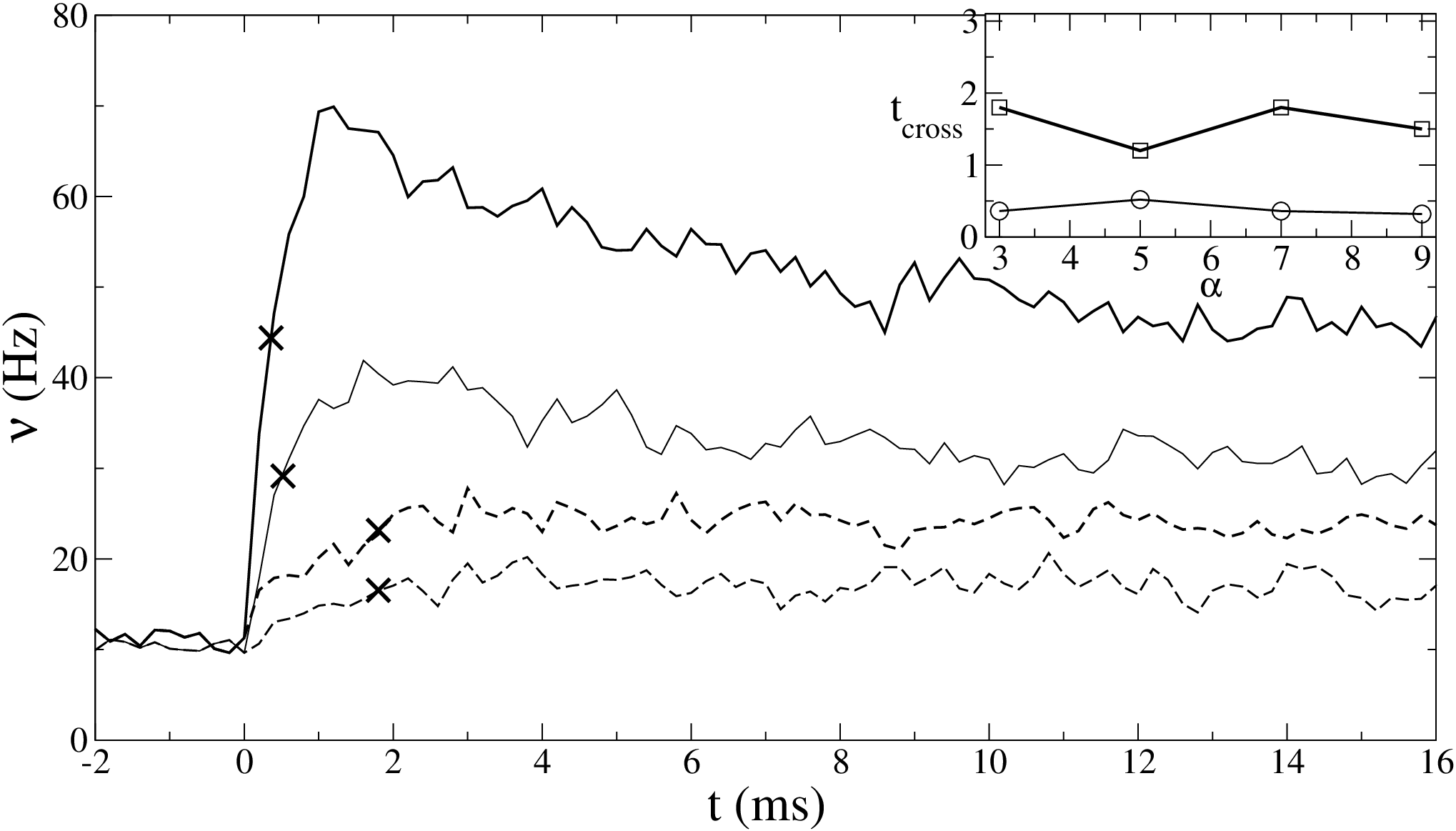}
\caption{\label{fig:histograma} 
Trial averages of the transient responses of a LIF neuron to changes
in the input correlations. Below $t=0$ the input has only
autocorrelations, described by $\mu=40s^{-1}$,
$\sigma^2_w=19.3s^{-1}$, $\alpha=0.56$, and $\tau_c=15ms$
(representing independent bursty input spike trains with, e.g.
$\nu_E=\nu_I=10Hz$, $N_E=10^4$, $N_I=2.10^3$, $F_E=4$, $F_I=1$,
$J_E=6.10^{-3}$, $J_I=2.8.10^{-2}$). {\bf Full lines:} quick responses
when $\alpha$ is suddenly changed at $t=0$ to $\alpha=7$ (upper curve)
and to $\alpha=3$ (bottom curve) and synchronization occurs in a
precise way ($\tau_c=1ms$). These two $\alpha$'s (corresponding to
different tone frequencies) can be obtained with $\rho_{EE}=0.34$ and
$\rho_{EE}=0.13$ respectively, and in both cases $f_{EE}=5.10^{-2}$,
$f_{II}=f_{EI}=f_{IE}=0$. {\bf Dashed lines:} the same as above but
$\tau_c=15ms$ after $t=0$, in agreement with \cite{deC+96}. The
responses are slower, but still fast in comparison with $\tau$. Other
parameters are as in Fig.  \ref{fig:grafica1}.  {\bf Inset:}
$t_{cross}$, time when the response hits for the first time the value
of the stationary rate (crosses in the main graph), as a function of
$\alpha$. Upper curve: $\tau_c=15ms$, bottom curve: $\tau_c=1ms$.}
\end{figure}

These predictions have been tested with numerical simulations in the
context of the experimental results found in \cite{deC+96}.  In this
experiment, neurons in primary auditory cortex (AI) are recorded under
stimulation by a pure tone. After the stimulus onset, a change in the
cross-correlogram is observed while the rate changes very little. The
results shown in Fig. \ref{fig:histograma} correspond to the response
of a LIF neuron integrating a current which emulates the activity in
AI. The input initially contains autocorrelations but not
cross-correlations and the output rate is low. When at $t=0$ a tone is
presented, there is a sudden increase in $\alpha$ (due to a
synchronization of a subpopulation in AI, which depends on the tone
frequency). The neuron responds by firing at a higher output rate. As
expected from Fig. \ref{fig:grafica2}, this final rate increases with
$\alpha$, but the velocity of the response is independent of it (see
inset Fig.\ref{fig:histograma}) This means that the reaction is
equally fast for any stimulation tone.  As a consequence of this
dynamics, the correlation coding present in AI is transformed into a
rate coding by the postsynaptic neuron.

In \cite{Sen+98} the same {\em synchrony reading} problem was discussed with
AI cells making depressing synapses with the reading neuron.
The authors show an example where a neuron with static synapses fails to
respond to the tone. We have checked that 
the results in Fig. \ref{fig:histograma} (dashed lines) hold for 
parameter values that can represent the experimental results.

Our results could be extended by including the effect of finite
synaptic time constants $\tau_s$; our work takes $\tau_s=0$ and thus
it is the zeroth order in an expansion in this parameter. Indeed, 
we have numerically checked that our conclusions hold qualitatively
if small, non-zero $\tau_s$ (e.g. $2$ ms) are considered.


\bibliography{bibliografia}

\end{document}